\documentclass[
10pt,twocolumn,notitlepage,oneside]{revtex4}
\usepackage{rotating}
\usepackage{amsfonts,longtable}

\usepackage{amsmath}
\usepackage{epsfig}
\usepackage{epstopdf}
\usepackage{amssymb}

\begin{document}

\title{Supermassive Black Holes and Nuclear Star Clusters: \\Connection with the Host Galaxy Kinematics and Color}

\author{\firstname{A.~V.}~\surname{Zasov}}
\email[]{zasov@sai.msu.ru}
\affiliation{Lomonosov Moscow State University, 
 Sternberg Astronomical Institute, Moscow, Russia}
%
\author{\firstname{A.~M.}~\surname{Cherepashchuk}}
\affiliation{Lomonosov Moscow State University, 
 Sternberg Astronomical Institute, Moscow, Russia}

\begin{abstract} We consider the relationship between the masses of the central objects in disky galaxies -- supermassive black holes (SMBHs) and nuclear star clusters (NCs) --  and various parameters of parent galaxies:
velocity of rotation $V_{(2)}$ at $R= 2$~kpc,  maximal velocity of  rotation
 $V_{\textrm{max}}$, the indicative dynamical mass $M_{25}$, the
total mass of the stellar population $M_{*}$, and the total color 
index ($B{-}V$) of galaxies.  The mass of nuclear clusters  $M_{\rm nc}$ correlates more 
closely with the kinematic parameters and total masses of galaxies  than the mass of central black holes $M_{\rm bh}$,
including   correlation with the velocity $V_{\textrm{max}}$, which characterizes 
the virial mass of the dark halo. On average, lenticular galaxies are 
 distinguished by higher masses $M_{\rm bh}$ compared to other types of
galaxies with similar characteristics. The dependence of masses of central objects
on the color index is bimodal: galaxies of the red group 
(red-sequence), which have ($B{-}V) > 0.6{-}0.7$, being mostly early-type 
galaxies, differ from bluer galaxies, 
by higher values of $M_{\rm nc}$ and $M_{\rm bh}$. The red-group galaxies have systematically 
higher $M_{\rm bh}$ values, even when the host-galaxy parameters are 
similar. In contrast, in the case of nuclear stellar clusters, the ``blue'' and 
``red'' galaxies form unified sequences.  The results agree with scenarios in 
which most red-group galaxies form as a result of the partial or complete
loss of interstellar gas in a stage of high nuclear activity which took place  in
galaxies whose central black-hole masses exceed $10^6 {-} 10^7 M_{\odot}$ 
(depending on the total mass of the galaxy). Most of galaxies 
with $M_{\rm bh}> 10^7 M_{\odot}$ are lenticular galaxies (types S0, 
E/S0) whose disks are essentially devoid of gas.
\end{abstract}

\maketitle

\section{INTRODUCTION}

There are two kinds of central massive objects which are observed at the dynamical 
centers of both disky and elliptical galaxies: supermassive black holes (SMBHs),
with masses of up to a few billion solar masses (see, e.g., the review
of Shankar [1]), and central (nuclear) star clusters (NCs), with
total masses of $10^5{-}10^7 M_{\odot}$ (see, e.g., the papers by
Wehner and Harris [2], Seth et al.~[3], and Ross et al.~[4]). As a rule,
the masses of nuclear star clusters are appreciably larger than the masses
of central black holes, except for the most massive galaxies, where the 
opposite is true [5, 6]. The genetic connection between NCs and SMBHs, their
formation mechanisms, and, especially, their evolution 
are poorly known and are subjects of active discussions (see, e.g., [6]).

The observations show that both NC and SMBH masses are correlated with the
properties of the host galaxies. The black-hole mass $M_{\rm bh}$ is most
closely related to the central stellar-velocity dispersion $\sigma$,
as well as the luminosity (mass) of the spheroidal system; i.e.,
the bulge for disk galaxies and the entire galaxy for elliptical systems
(see, e.g., [7] and references therein). On the contrary, the NC masses 
$M_{\rm nc}$ are more closely related to the stellar components of their
galaxies, and, unlike the SMBHs, they are often observed in 
late-type galaxies where a bulge is nearly absent [8, 9]. Like the SMBH
mass, the mass of NC is correlated with the central velocity dispersion, 
though not so closely [10--12]. However, the NCs and SMBHs do not form 
a unified sequence: the slopes of the logarithmic dependencies of $M_{\rm nc}$ 
on $\sigma$ (as well as their dependencies on the luminosity 
and mass of the  stellar components) are gentler than the 
corresponding slopes for $M_{\rm bh}$, suggesting different formation 
mechanisms and evolutionary trends for these objects [12].

When comparing SMBHs and NCs, we should bear in mind that most galaxies 
with reliable $M_{\rm bh}$ estimates are massive ellipticals, in which 
nuclear star clusters are rarely observed, while their masses have 
been measured mostly in low-mass, late-type (Sd--Sm) galaxies, in which
the velocity dispersion $\sigma$ and bulge mass are small, so that
few estimates of the SMBH masses are available for them. This makes difficult to compare these  two types of central objects. A correlation between $ M_{\rm bh}$ and
$\sigma$ is clearly expressed for early-type (E--S0) galaxies, but 
is smeared out and almost vanishes completely for late-type galaxies 
(which have low values of $M_{\rm bh}$, as a rule). 

As it was pointed out 
by Erwin and Gadotti [9], the stellar-velocity dispersion along the line 
of sight is not a very good parameter for comparing galaxies whose  
spherical component is small. This is due to the fact that it is hard to
divide the measured stellar-velocity dispersion in galaxies with small
bulges into separate components belonging to the nuclear cluster, the
bulge, or the inner disk. Moreover, an estimated velocity dispersion 
generally depends on the radius of the region in which it is determined. 
It is not unambiguously related to either the bulge mass or the total 
mass within a given radius: the mass of a stellar system estimated 
from its velocity dispersion is always model-dependent.

In the case of disk galaxies, another approach is possible: one may compare the masses 
of the central objects with direct estimates of the disk rotation velocities,
instead of the velocity dispersion. 
One can use the maximum rotation velocities, which, as a rule, are close to 
the ``asymptotic'' circular velocities at large distances from the center,  or the velocities at a specified distance in the inner 
region of the disk, if our interest is in the density of the central region 
of the galaxy [11].  A shortcoming of this approach is the lower (on 
average) accuracy of the estimated rotation velocity, compared to estimates of
the velocity dispersion, especially given the dependence of the velocity 
on the adopted disk inclination and the possible presence of non-circular 
or non-planar motions, which are often observed in the 
central regions of galaxies. 

The first attempts to identify kinematic 
features in the central kiloparsec of galaxies with active nuclei were 
undertaken by Afanasyev [13] in the 1980s. Zasov et al. [11, 14] applied 
this approach to SMBHs, for which the estimates of $M_{\rm bh}$ have mostly 
been obtained via reverberation mapping or direct kinematic measurements. In particular, it was shown in [11] that the mass  $M_{\rm bh}$ 
 is correlated 
with the disk rotation velocity $V_{(1)}$ at a distance of $\sim 1$ kpc, i.e., with the density of the central region, which agrees
well with the scenario where the bulge and the SMBH are formed as a result of
a monolythic collapse of the central region of the protogalaxy. In the present work we use the rotational velocity at $R= 2$~kpc, $V_{(2)}$,  because many galaxies
do not have reliable estimates of $V_{(1)}$, corresponding to a distance 
of 1 kpc, due to the limited linear resolution of their rotation curves. 
However, as we found, a transiton from $V_{(1)}$ to $V_{(2)}$ does 
not qualitatively change the form of the dependencies, however it increases their
statistical significance.

The dark halo could theoretically play a large 
role in SMBH formation, and in determining the final black hole mass [15--21].  The total (virial) mass of the dark halo is determined by the  circular velocity at the virial radius, which is close to  the  observed maximum or ``asymptotic'' rotation velocities $V_{max}$  
of galaxies at large $R$ [22, 23].  So the relationship between the masses of the central objects, $M_{\rm bh}$ or $M_{\rm nc}$, and $V_{max}$  are of special interest.

The connection between the SMBH or NC masses and galactic circular
velocities has been considered in a number of studies, but the results are 
contradictory. It is obvious, that the most massive black holes are observed 
in high-luminosity galaxies, which have
high rotation velocities. However, the tight correlation  $M_{\rm bh}$ -- $V_{\textrm{max}}$ for disk galaxies found in some studies [21, 24] is probably the result of using 
indirect estimates of $M_{\rm bh}$, based on 
their statistical dependencies on the central stellar-velocity dispersion.  
In the paper of Beifiori at al. [25], where the maximum rotation 
velocity was estimated from the width of the HI line, the correlation between $M_{\rm bh}$ 
and $V_{\textrm{max}}$ for disk galaxies turned out to be very weak, and 
became significant only after the addition of elliptical galaxies with 
model estimates of their circular velocities (see Fig.~13 of [25]). A weak 
correlation between these parameters was also noted for early-type disk galaxies
in [14], where real rotation curves of galaxies were used.  According to 
Kormendy and Bender [8], $M_{\rm bh}$ is only correlated with the parameters 
of ``classical'' bulges, and does not depend on the parameters of the disk 
(which can even be absent); therefore, a correlation between $M_{\rm bh}$ 
and $V_{\textrm{max}}$ is clearly expressed only for galaxies with massive 
bulges, probably as a result of a link between the bulge parameters and 
the rotation velocity of the galaxy at large $R$ (the ``baryon--dark matter 
conspiracy''). However,   
Volonteri et al. [17] came to conclusion that there is no discrepancy between the
theoretically expected correlation between  $M_{\rm bh}$ and  $V_{\textrm{max}}$ and observations.

In the current study, we have analyzed the connection between the masses
of nuclear star clusters and central supermassive black holes with the 
kinematic parameters and integrated $B{-}V$ color indices of their host 
disky galaxies. We don't consider purely elliptical galaxies without disks here.

\section{GALAXY SAMPLE AND DATA SOURCES}

Our sample of galaxies with known NC masses is based on the list of Seth
et al. [3], to which we have added a few objects from the  lists of galaxies with the most confident NC mass estimates compiled by Erwin and Gadotti [9], Graham and Spitler [5] and Graham
[26]. The sample of galaxies with the SMBH masses is based on the 
catalog of Graham [26], supplemented by data from [7, 27], as well as a 
few galaxies we had observed earlier [11, 28]. Since the number of galaxies 
with $\log M_{\rm bh}< 6$ is very small, we added a few objects with 
low mass black holes, although for some of them only upper limits of $M_{bh}$ are known (IC~342~[29] , NGC~404~[30], NGC~598~[31], and NGC~4395~[32],  and several galaxies from the list of Neumayer et al [6]). Our list also includes NGC 1277, which possesses the most massive SMBH [32]. We adopted the inclination angle  $i= 52^{\circ}$ [33] for this galaxy to get  $V_{\textrm{max}}$ from the line-of-sight velocity curve given in the cited paper.

The characteristic errors of the $M_{\rm bh}$ estimates are about a factor 
of two~[7, 27]; the errors are smaller for the most massive and larger for 
less massive black holes. The accuracy of the NC masses values is 
generally also close to a factor of two [5]. The adopted distances to 
the galaxies correspond to $H_0 = 75$~km\,s$^{-1}$Mpc$^{-1}$. The results of a comparison of the SMBHs and NCs will depend only weakly 
on this parameter. For nearby galaxies ($V_r \le 800$~km/s), we adopted the 
same distances as in the papers where the rotation velocities were estimated
(the exception is NGC~4258, for which Sofue et al. [34] apparently used
a strongly overestimated distance of 6.6~Mpc). The distance to 
the Virgo cluster was taken to be 16~Mpc. In a few cases, where the 
accepted distance was substantially different from the distance given in 
the papers presenting the central-object mass estimates, the appropriate
corrections were applied.

\begin{longtable}[h!]{llccccl}
\caption{Galaxies with known rotation velocities}\\
\hline 
\hline
{Galaxy}&{Type}&$D$&$\log M_{\rm nc}$&$\log M_{\rm bh}$&$V_{max}$& Source\\
            &          &Mpc&$10^6~M_\odot$ &$10^6~M_\odot$  &km/s        & \\
\endfirsthead
\caption{Continued}\label{Tabl1} \\
\hline 
{Galaxy}&{Type}&$D$&$\log M_{\rm nc}$&$\log M_{\rm bh}$&$V_{max}$& Source\\
            &          &Mpc&$10^6~M_\odot$ &$10^6~M_\odot$  &km/s        & \\
\hline
\endhead
\hline
\endfoot

\hline
Circinus    &   Sb  &   4   &   \multicolumn{1}{c}{--}   &   6.2 &   150 &   [36]    \\
IC 0342  &   Scd &   3.9 &   7.1 &  $<$6  &   193 &   [34]    \\
IC 2560  &   SBb &   40.7    &   \multicolumn{1}{c}{--}   &   6.6 &   ~~196$^*$    &   [33]    \\
MW  &   SBbc    &     \multicolumn{1}{c}{--} &   7.5 &   6.6 &   230 &   [37]    \\
N\,0224 &   Sb  &   0.77    &   7.5 &   8.2 &   250 &   [38]    \\
N\,0253 &   SBc &   3.5 &   \multicolumn{1}{c}{--}   &   7   &   210 &   [34]    \\
N\,0289 &   SBbc    &   21.4    &   7.9 &   \multicolumn{1}{c}{--}   &   170 &   [39]    \\
N\,0300 &   Scd &   1.8 &   6   &  $<$3  &   \phantom{0}90  &   [40]    \\
N\,0428 &   SABm    &   15.9    &   6.5 &  $<$4.5    &   110 &   [28]    \\
N\,0450 &   SABc    &   24.3    &   6.1 &   \multicolumn{1}{c}{--}   &   130 &   [41]    \\
N\,0524 &   S0/a    &   35  &   \multicolumn{1}{c}{--}   &   8.92    &   360 &   [42]    \\
N\,0598 &   Sc  &   0.84    &   6.3 &  $<$3.2    &   130 &   [43]    \\
N\,1023 &   E/SO    &   11.7    &   6.4 &   7.64    & 205 &   [44]    \\
N\,1042 &   SABc    &   18.2    &   6.5 &  $<$4.4    &   \phantom{0}52  &   [45]    \\
N\,1068 &   Sb  &   15  &   \multicolumn{1}{c}{--}   &   6.9 &   240 &   [46] \\
N\,1277 &   S0  &   73  &   \multicolumn{1}{c}{--}   &   10.2    &   350 &   [32]    \\
N\,1300 &   SBbc    &   20  &   7.9 &   7.9 &   210 &   [47]    \\
N\,1325 &   SBbc    &   20.1    &   7.1 &   \multicolumn{1}{c}{--}   &   184 &   [48]    \\
N\,1345 &   Sc  &   19.4    &   6.1 &   \multicolumn{1}{c}{--}   &   100 &   [45]    \\
N\,1385 &   Sc  &   18.7    &   6.4 &   \multicolumn{1}{c}{--}   &   130 &  [49]  \\
N\,1705 &   S0  &   6.2 &   5   &   \multicolumn{1}{c}{--}   &   \phantom{0}70  &   [50]    \\
N\,2139 &   Sc  &   22.2    &       &  $<$5.2    &   136 &   [33]    \\
N\,2549 &   S0  &   16  &   7   &   7.26    &   160 &   [44]    \\
N\,2552 &   SABm    &   10.1    &   5.8 &   \multicolumn{1}{c}{--}   &   \phantom{0}90  &   [51]    \\
N\,2778 &   E/S0    &   27  &   \multicolumn{1}{c}{--}   &   7.26    &   135 &   [52]    \\
N\,2787 &   SB0 &   11.5    &   \multicolumn{1}{c}{--}   &   7.8 &   200 &   [28]    \\
N\,2964 &   Sbc &   17  &   7.8 &   \multicolumn{1}{c}{--}   &   220 &   [53]    \\
N\,3079 &   Sb  &   16  &   \multicolumn{1}{c}{--}   &   6.28    &   230 &   [34]    \\
N\,3115 &   S0  &   10.3    &   \multicolumn{1}{c}{--}   &   8.96    &   240 &   [54]    \\
N\,3227 &   S0  &   20.3    &   \multicolumn{1}{c}{--}   &   7.15    &   250 &   [55]    \\
N\,3245 &   S0  &   18  &   \multicolumn{1}{c}{--}   &   8.25    &   250 &   [28]    \\
N\,3346 &   SBc &   15.6    &   6.1 &   \multicolumn{1}{c}{--}   &   110 &   [51]    \\
N\,3368 &   SBab    &   10.1    &   \multicolumn{1}{c}{--}   &   6.86    &   210 &   [56]    \\
N\,3384 &   S0  &   11.6    &   7.3 &   7.23    &   160 &   [57]    \\
N\,3414 &   S0  &   18  &   \multicolumn{1}{c}{--}   &   8.4 &   145 &   [58]    \\
N\,3423 &   Sc  &   11.9    &   \multicolumn{1}{c}{--}   &  $<$5.2    &   127 &   [33]    \\
N\,3489 &   S0  &   11.7    &   \multicolumn{1}{c}{--}   &   6.76    &   \phantom{0}54  &   [33]    \\
N\,3501 &   Sc  &   14  &   5.9 &       &   126 &   [59]    \\
N\,3516 &   S0  &   36  &   \multicolumn{1}{c}{--}   &   7.6 &   200 &   [28]    \\
N\,3585 &   SB0 &   18  &   6.5 &   8.49    &   140 &   [57, 60] \\
N\,3607 &   S0  &   22  &   \multicolumn{1}{c}{--}   &   7.94    &   250 &   [61]    \\
N\,3621 &   Sd  &   6.6 &   7   &   \multicolumn{1}{c}{--}   &   150 &   [62]    \\
N\,3949 &   Sbc &   11.4    &   6.9 &   \multicolumn{1}{c}{--}   &   170 &   [39]    \\
N\,3998 &   S0  &   15  &   \multicolumn{1}{c}{--}   &   8.34    &   305 &   [57]    \\
N\,4026 &   S0  &   13.2    &   6.7 &   8.26    &   200 &   [57]    \\
N\,4027 &   SBd &   20.1    &   5.9 &   \multicolumn{1}{c}{--}   &   110 &   [63]    \\
N\,4144 &   SABc    &   4.3 &   4.8 &       &   \phantom{0}80  &   [64]    \\
N\,4183 &   Sc  &   12.9    &   5.9 &       &       &   [65]    \\
N\,4206 &   Sbc &   16  &   6.8 &   \multicolumn{1}{c}{--}   &   100 &   [66]    \\
N\,4244 &   Sc  &   5.2 &   6.5 &   \multicolumn{1}{c}{--}   &   100 &   [67]    \\
N\,4258 &   SBbc    &   6.6 &   \multicolumn{1}{c}{--}   &   7.59    &   213 &   [34]    \\
N\,4395 &   Sm  &   4.6 &   6.1 &   5.6 &   90  &   [51]    \\
N\,4459 &   S0  &   16  &   \multicolumn{1}{c}{--}   &   7.85    &   300 &   [68]    \\
N\,4496 &   SBd &   16  &   5.7 &   \multicolumn{1}{c}{--}   &   110 &   [59]    \\
N\,4564 &   S0  &   16  &   \multicolumn{1}{c}{--}   &   7.78    &   150 &   [69 \\
N\,4625 &   SABm    &   9.5 &   5.6 &   \multicolumn{1}{c}{--}   &   \phantom{0}37  &   [70]    \\
N\,4945 &   SBcd    &   3.8 &   \multicolumn{1}{c}{--}   &   6.15    &   170 &   [70]    \\
N\,5023 &   Sc  &   4.8 &   5.3 &   \multicolumn{1}{c}{--}   &   \phantom{0}83  &   [71]    \\
N\,5128 &   S0  &   3.8 &   \multicolumn{1}{c}{--}   &   7.65    &   230 &   [72]    \\
N\,5377 &   Sa       &   25.3    &   8.6 &   \multicolumn{1}{c}{--}   &   220 &   [73]    \\
N\,5584 &   SABc    &   21.6  &   5.1 &   \multicolumn{1}{c}{--}   &   100 &   [74]    \\
N\,5585 &   SABc    &   10.5    &   5.8 &       &   100 &   [51]    \\
N\,5669 &   SABc    &   18.5    &   6.5 &   \multicolumn{1}{c}{--}   &   \phantom{0}98  &   [75]    \\
N\,5678 &   SABb    &   27.2    &   8.1 &   \multicolumn{1}{c}{--}   &   187 &   [59]    \\
N\,5806 &   Sb  &   18  &   8.1 &   \multicolumn{1}{c}{--}   &   200 &   [39]    \\
N\,5879 &   Sbc &   12.3  &   7.2 &   \multicolumn{1}{c}{--}   &   140 &   [76]    \\
N\,5964 &   SBcd    &   19.9    &   6.4 &   \multicolumn{1}{c}{--}   &   112 &   [35]    \\
N\,6239 &   SBb &   14.6    &   6.7 &   \multicolumn{1}{c}{--}   &   \phantom{0}84  &   [59]    \\
N\,6509 &   SABc    &   26  &   6.4 &   \multicolumn{1}{c}{--}   &   218 &   [75]    \\
N\,6946 &   Scd &   5.9 &   7.9 &   \multicolumn{1}{c}{--}   &   200 &   [62]    \\
N\,6951 &   SABb    &   21.9    &   8.2 &   \multicolumn{1}{c}{--}   &   230 &   [34]    \\
N\,7418 &   Sc  &   19.3    &   7.8 &  $<$5.2    &   155 &   [33]    \\
N\,7424 &   Sc  &   12.3    &   6.1 &  $<$5.2    &   \phantom{0}82  &   [33]    \\
N\,7457 &   E-S0    &   14.6    &   7   &   6.54    &   130 &   [11]    \\
N\,7469 &   Sa  &   68  &   \multicolumn{1}{c}{--}   &   7.1 &   170 &   [11]    \\
N\,7793 &   Scd &   3.4 &   6.9 &  $<$3.7    &   115 &   [77]    \\
U\,03826    &   SABc    &   24  &   5.6 &   \multicolumn{1}{c}{--}   &   \phantom{0}70  &   [76]    \\
U\,04499    &   Sd  &   13  &   4.9 &   \multicolumn{1}{c}{--}   &   \phantom{0}60  &   [71]    \\
U\,06983    &   Sd  &   15.3    &   5.5 &   \multicolumn{1}{c}{--}   &   107 &   [65]    \\
U\,08823   &   S0  &   126 &   \multicolumn{1}{c}{--}   &   7.54    &   230 &   [11]    \\
U\,12732    &   SABm    &   13.2    &   5.8 &   \multicolumn{1}{c}{--}   &   \phantom{0}65  &   [71]    \\
VCC1242 &   S0  &   16  &   7.1 &   \multicolumn{1}{c}{--}   &   120 &   [78]    \\
\hline
\hline
\end{longtable}

The velocity $V_{\textrm{max}}$ used here corresponds to the plateau or
the maximum in the rotation curve, if it is located further than 
2~kpc from the center to avoid the cases where it is related to a bulge. The galaxy inclination $i$ was taken
from the original paper, or, if not presented there, from the Hyperleda 
database [33]. An exception is the NGC~5964, for which we adopted 
$i=42^{\circ}$ instead of the underestimated value $i=20^{\circ}$ used
by Fathi et al. [35]. No corrections for asymmetric drift were introduced for 
rotation curves obtained from absorption lines (if it was not given 
in the original paper).  However, we excluded from consideration those galaxies where 
the velocity dispersions outside the bulge is comparable with  $V_{\textrm{max}}$, making the correction 
for asymmetric drift substantial. In order not to exclude from consideration those galaxies which
have only upper limits for the SMBH mass, we took for them, if necesary,  $V_{\textrm{max}}$ estimates obtained from the width of the HI line and given in Hyperleda [33]. Conditionally
we adopted the accuracy of the rotation velocities to be ${\pm} 20\%$ (or 
$\Delta \log V\approx 0.08)$. The indicative mass within the 
25~mag/arcsec$^2$ isophote, which is proportional to $V_{max}^2$, typically has 
an uncertainty of a factor of $\sim 1.5$, which corresponds to 
$\Delta \log M_i \approx 0.18$. The central velocity dispersions have higher accuracies
than the rotation velocities  (about 10$\%$)~[7].

The data for galaxies with the $V_{\textrm{max}}$ estimates are given in Table 1. This table also presents the morphological types of the galaxies, 
the adopted distances and references.

\section{DEPENDENCE OF THE MASS OF CENTRAL OBJECTS ON THE ROTATION
VELOCITY AND HOST-GALAXY MASS}

Figures 1a,b show the velocity of rotation $V_{(2)}$ at $R=2$~kpc 
plotted against the central-object mass for NCs and SMBHs, respectively.
The velocity $V_{(2)}$ characterizes the disk angular velocity in the
inner part of a galaxy, being the indicative of the mean density 
within 2~kpc. The filled triangles denote galaxies for which only upper 
limits for $M_{\rm bh}$ are known. The galaxies of our sample with the $M_{nc}$ and $M_{bh}$ values were divided into two 
nearly equial groups: early types (hollow symbols) and late 
types (filled symbols).  We refer to the early types the lenticular (S0 and E/S0) galaxies with known $M_{bh}$ , while
 in the case of the galaxies with known $M_{nc}$ we  also assign  to the early types Sa-Sb galaxies, 
since the number of S0 galaxies among them is very small. The letter 
``e'' denotes galaxies whose disks are seen almost edge-on ($i > 
80^{\circ}$~[33]), and ``B'' denotes those with a bar (SB, SB0). Due to 
projection effects and the inner  absorption, the rotation 
velocities in the inner regions of galaxies seen almost edge-on may be 
underestimated [79], but this effect is not appreciable in either Fig.~1 
or the other diagrams considered below (though the positions of edge-on galaxies  are only 
marked in Fig. 1a).  The same is true for SB galaxies: we didn't find any  systematic
differences between galaxies with and without bars in the diagrams. 
The straight regression line in Fig. 1a and the other diagrams below is a bi-sector line obtained by a least-squares fit. Regression line in Fig 1b is the same as in Fig. 1a, being moved there for comparison. 		
\begin{figure*}
\includegraphics[width=7.5cm,keepaspectratio]{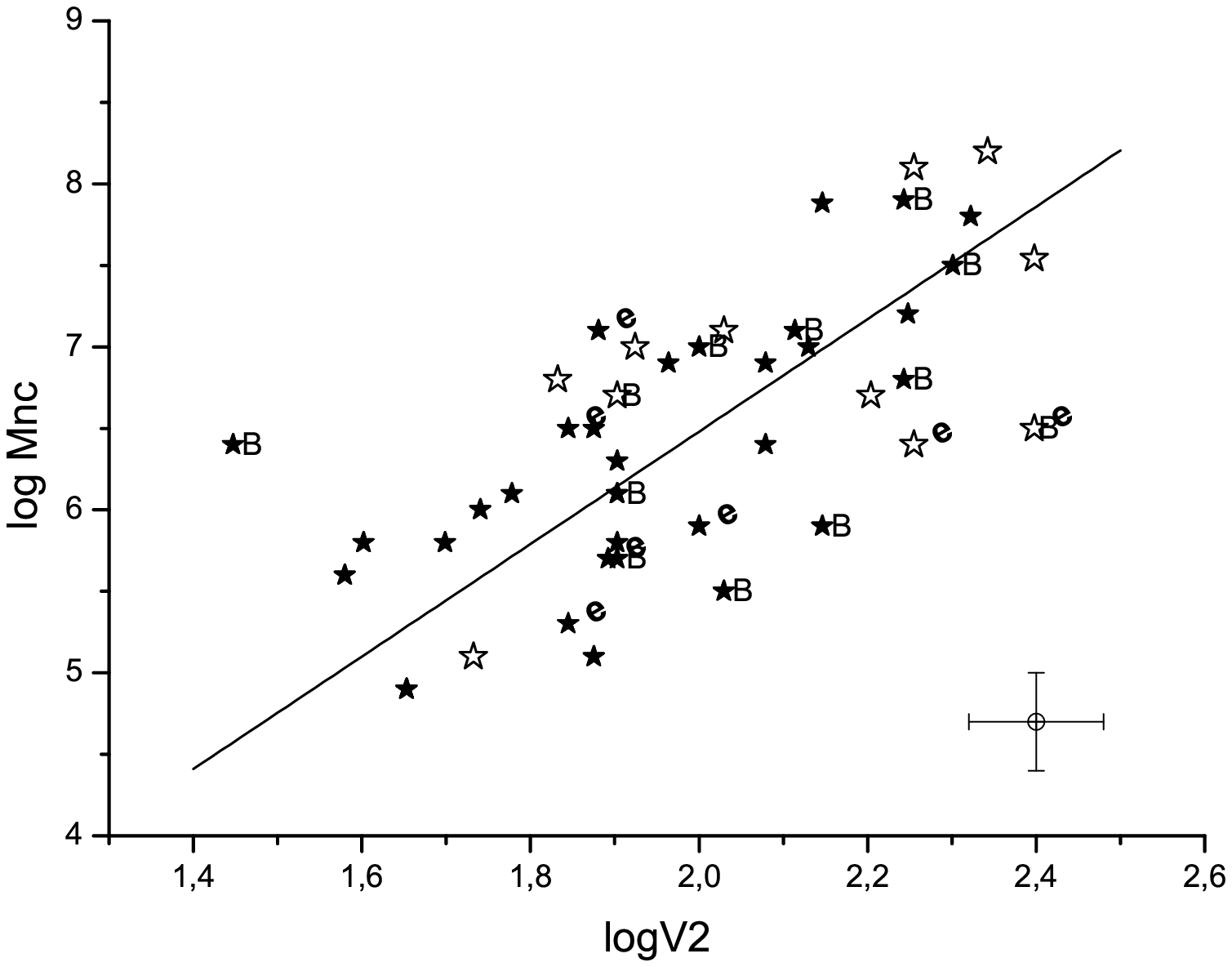}
\includegraphics[width=7.5cm,keepaspectratio]{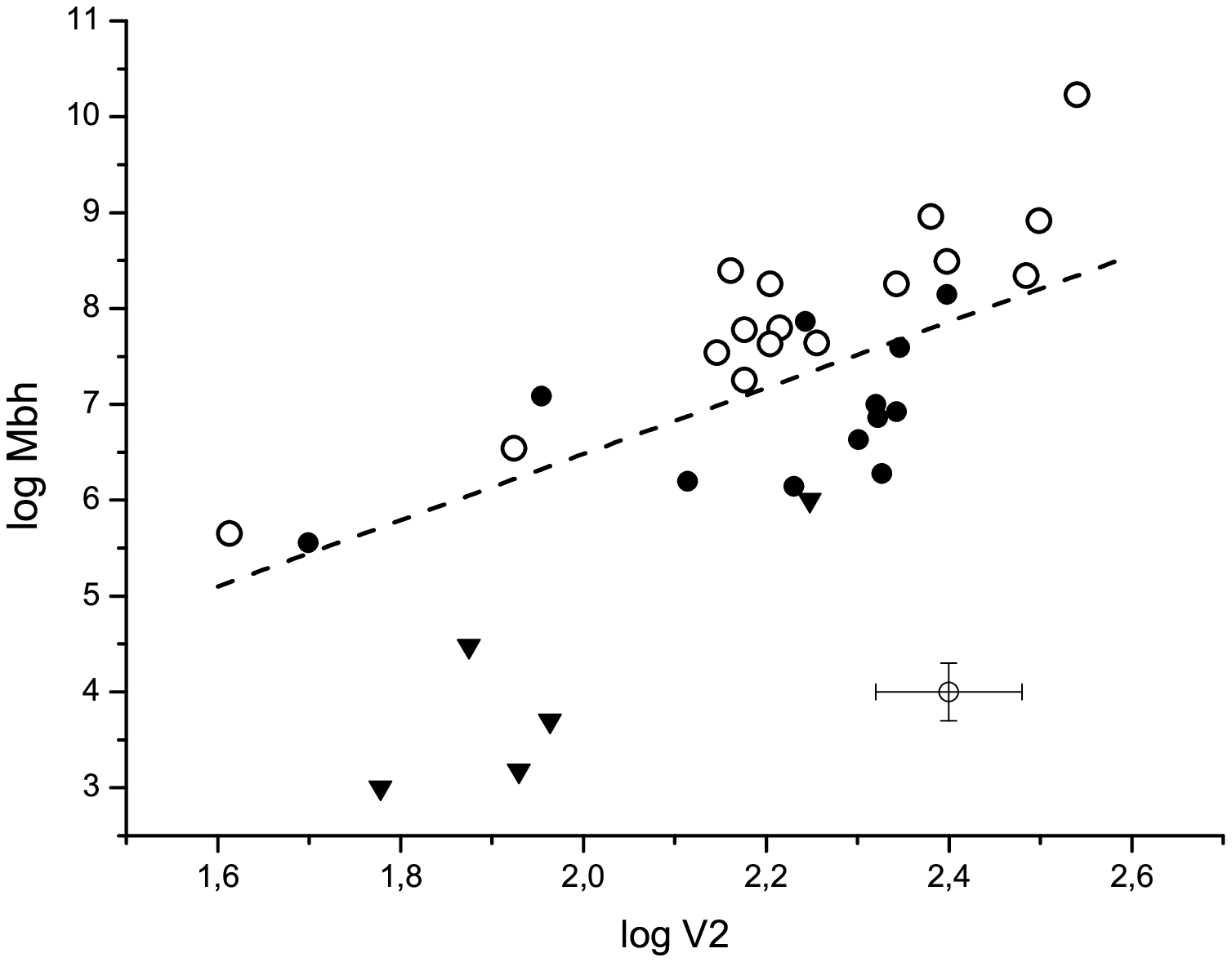}
\caption{ Dependence of (a) $M_{\rm nc}$ and (b) $M_{\rm bh}$ on the
velocity of rotation at $R = 2$~kpc,  $V_{(2)}$. The hollow symbols show
data for early-type galaxies (lenticular galaxies in the case (b)).
The filled triangles are galaxies with only upper limits for 
$M_{\rm bh}$. In (a), the symbols ``e'' and ``B'' mark galaxies 
observed edge-on and containing a bar, respectively. Their positions do 
not systematically differ from those of other galaxies. The solid line 
in (a)
is reproduced as the dashed line in graph (b).}
\label{fig1}
\end{figure*}


A comparison of Figs. 1a and 1b shows that the masses of both NCs and SMBHs 
 correlate with the angular velocities of 
central regions, but this correlation is shallower for black holes,
mainly owing to the galaxies with the upper limits for 
$M_{\rm bh}$. Early-type and late-type galaxies are not separated in 
the NC plot, however they occupy distinct regions in the SMBH plot: the 
lenticular galaxies (hollow symbols) possess more massive black holes
than late-type galaxies with the same velocities $V_{(2)}$. The 
dependence is shallower for NCs as well as for SMBHs.
\begin{figure*}
\includegraphics[width=7.5cm,keepaspectratio]{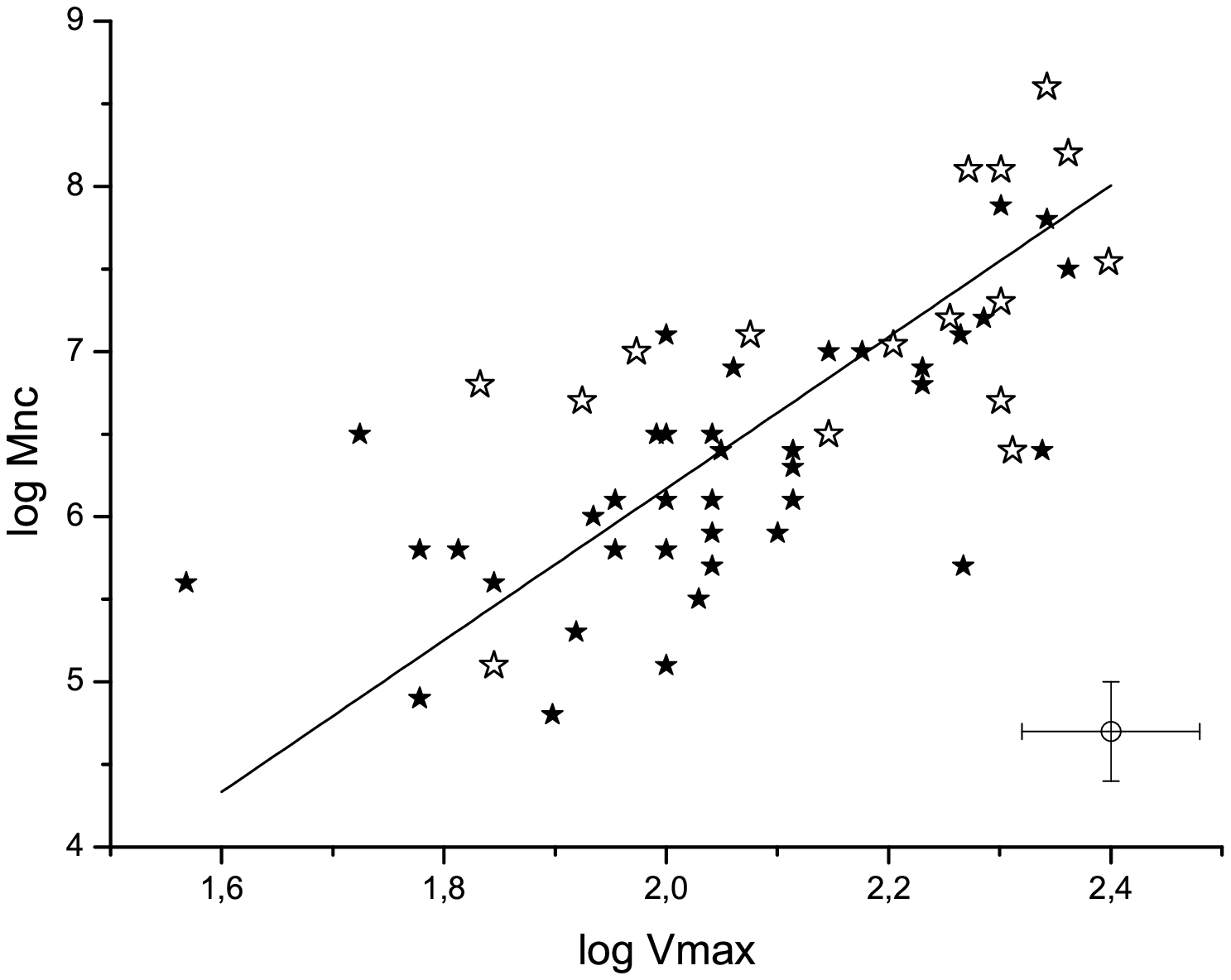}
\includegraphics[width=7.5cm,keepaspectratio]{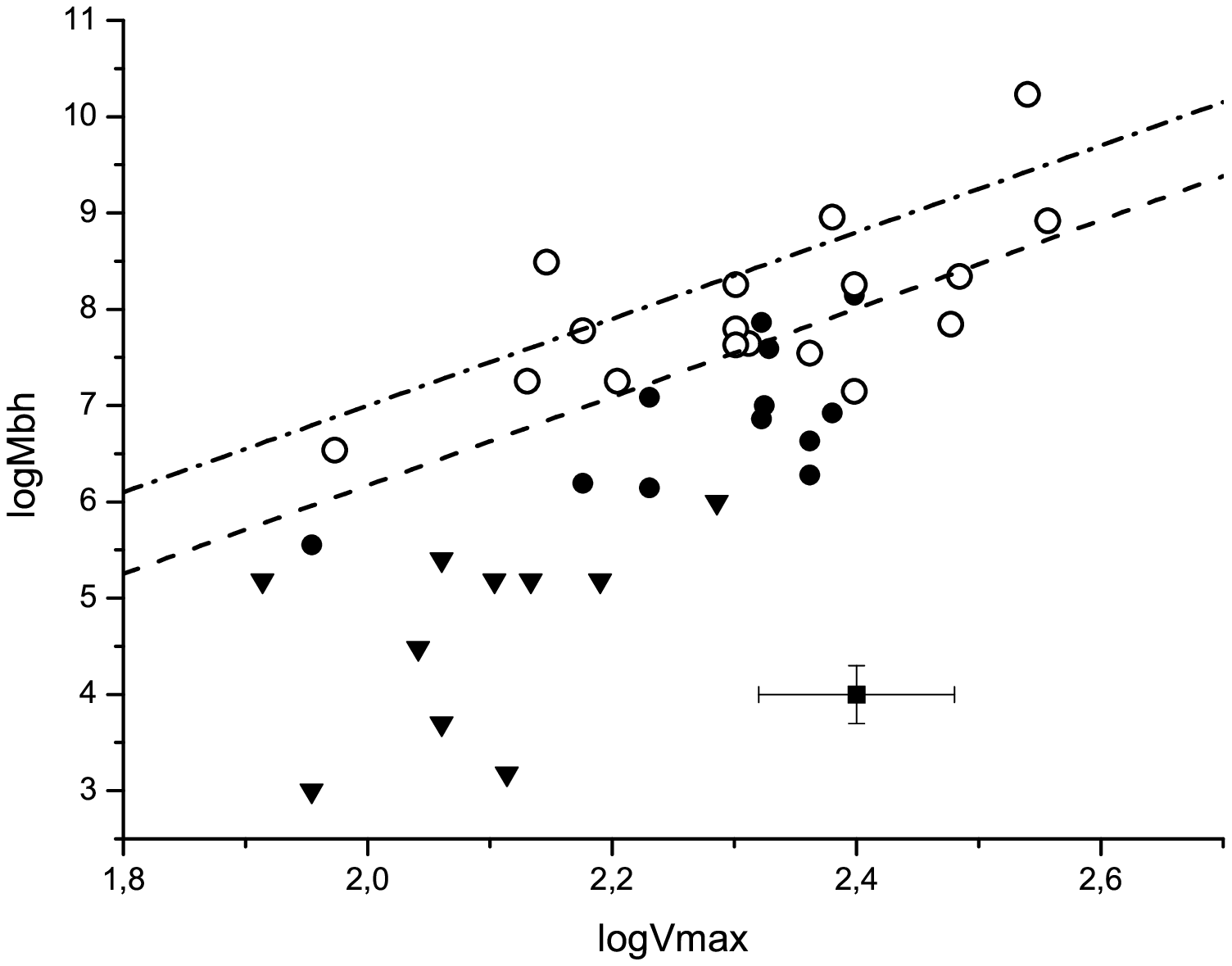}
\caption{ Dependence of (a) $M_{\rm nc}$ and (b) $M_{\rm bh}$ on the 
maximum (asymptotic) disk velocity. The notation is the same as in 
Fig. 1. The dot-dashed line in (b) shows the dependence $M_{\rm bh}\sim
V_{\textrm{max}}^{4.5}$, predicted by numerical models (see the text); 
its shift along the vertical axis is arbitrary.}
\label{fig2}
\end{figure*}


\begin{figure*}
\includegraphics[width=7.5cm,keepaspectratio]{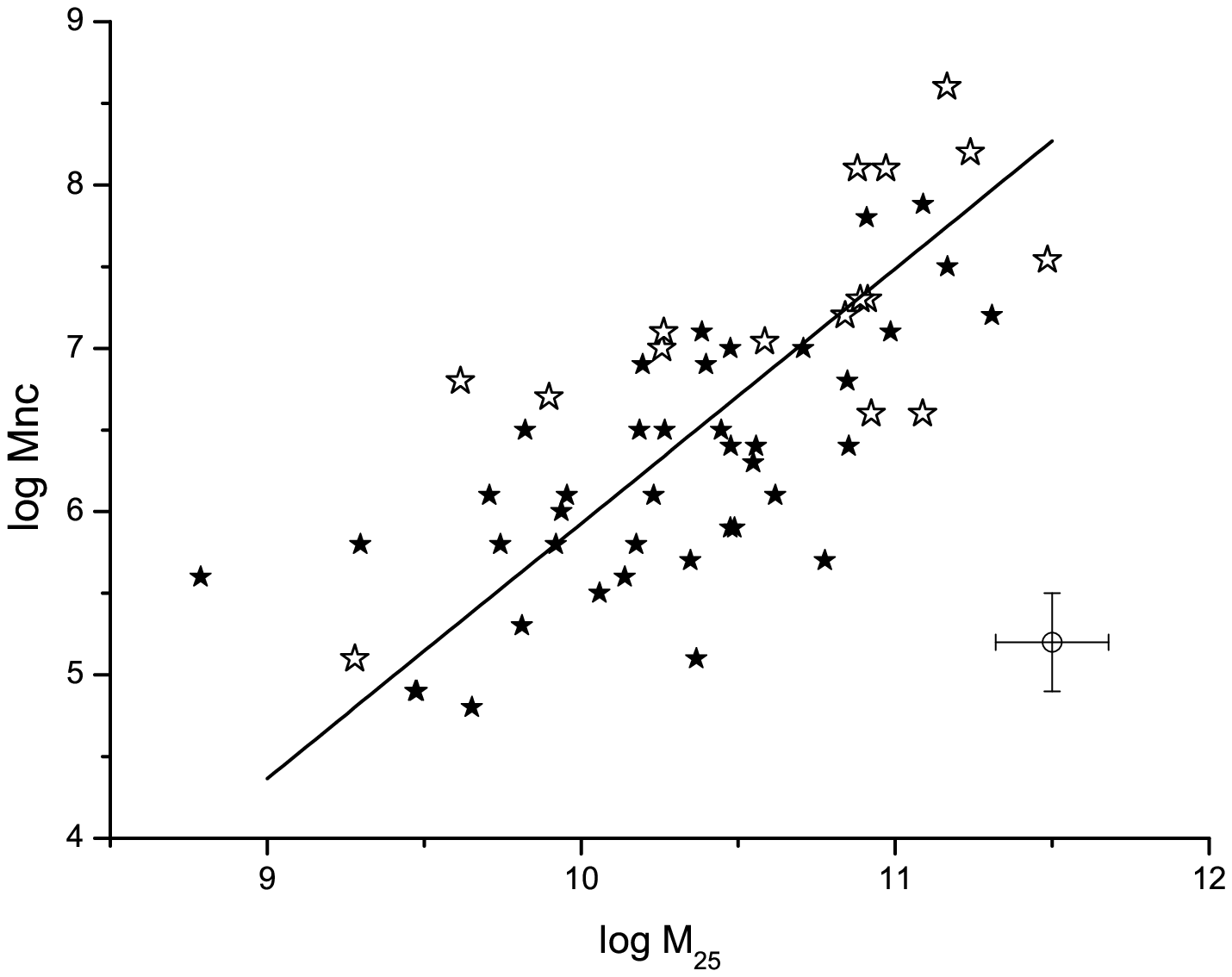} 
\includegraphics[width=7.0cm,keepaspectratio]{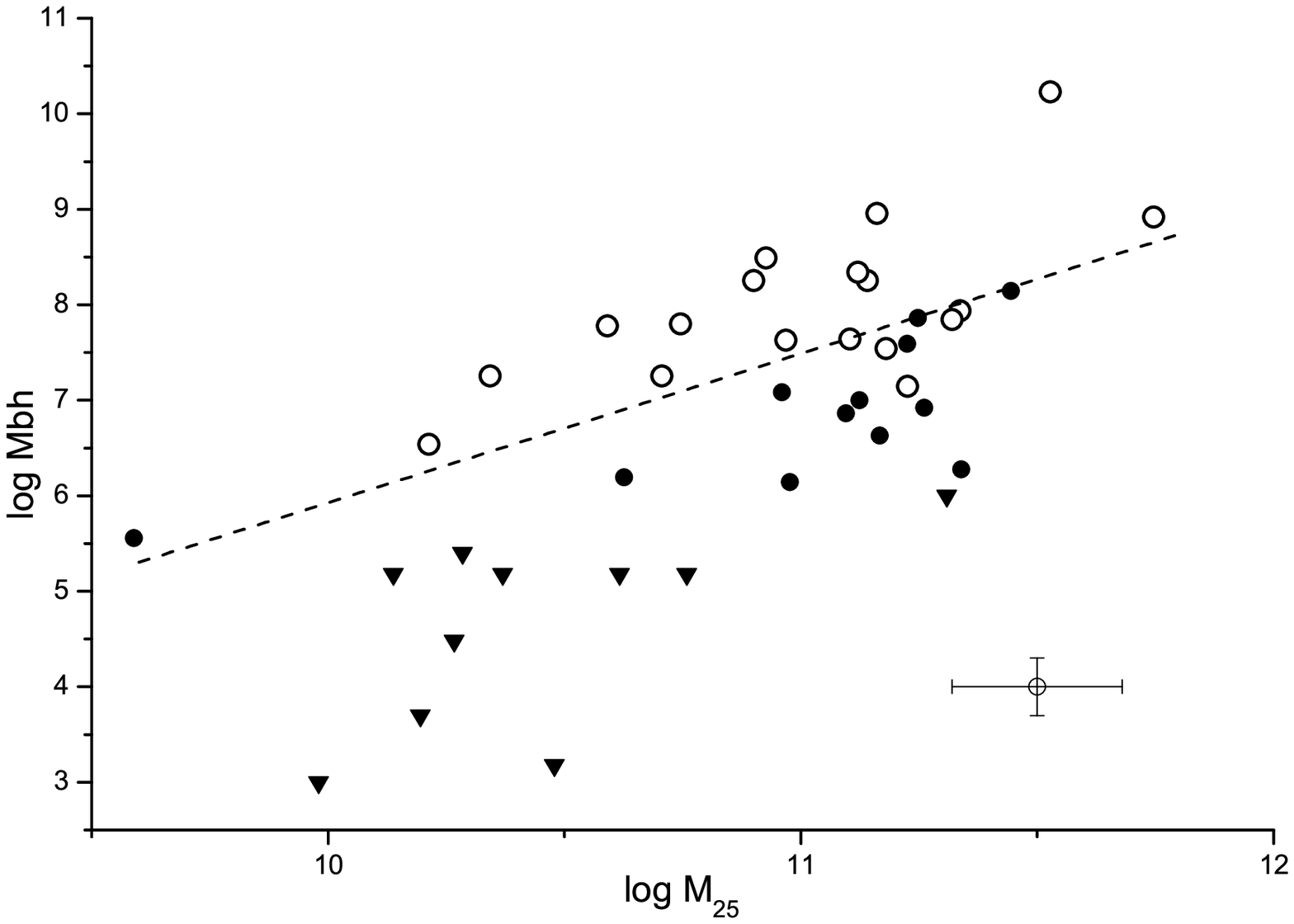}
\caption{Dependence of (a) $M_{\rm nc}$ and (b) $M_{\rm bh}$ on the
dynamic mass $M_{25}$ within the optical radius. The notation is
the same as in Figs. 1a,b.}
\label{fig3}
\end{figure*}

\begin{figure*}
\includegraphics[width=7.5cm,keepaspectratio]{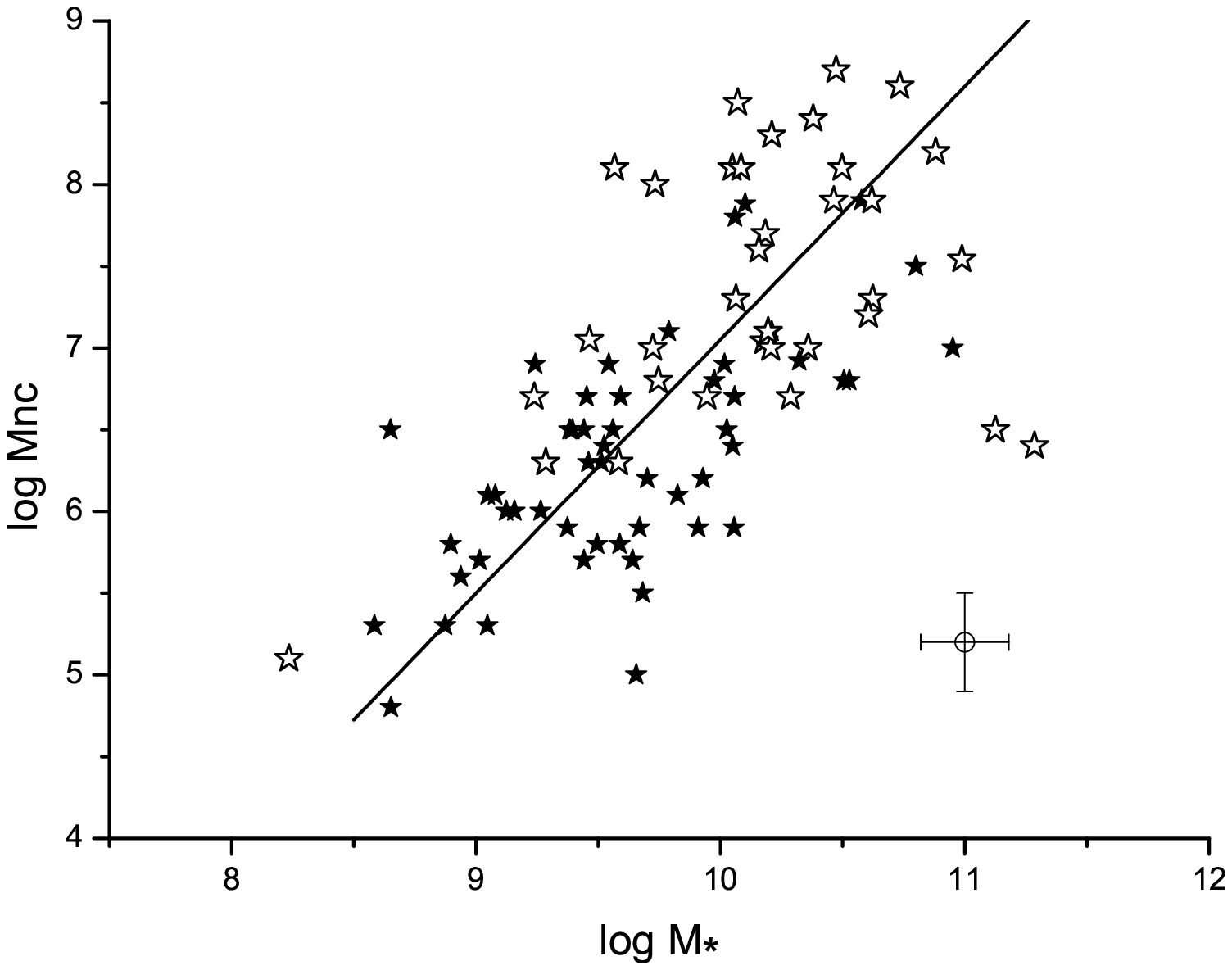}
\includegraphics[width=7.5cm,keepaspectratio]{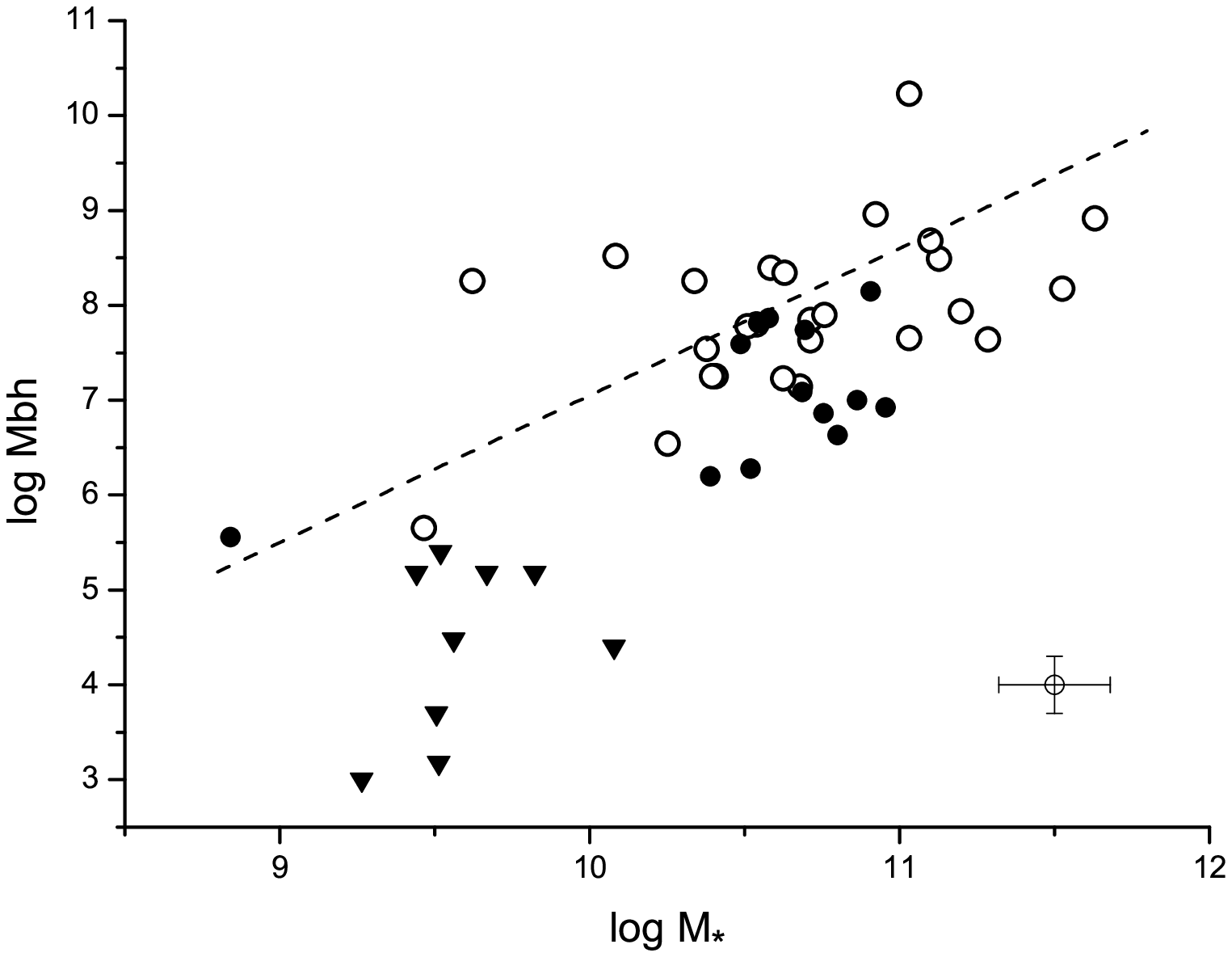}
\caption{Dependence of (a) $M_{\rm nc}$ and (b) $M_{\rm bh}$ on the
mass of the stellar population of parent galaxy. The notation is 
the same as in Figs. 1a,b.}
\label{fig4}
\end{figure*}


Similar conclusions follow from a comparison of the mass of  central
objects with the maximum rotation velocity $V_{\textrm{max}}$ of parent galaxy
(Fig.~2) and with the total (indicative) mass of a galaxy within the optical
diameter $D_{25}$, defied as $M_{25} =
V_{\textrm{max}}^2D_{25}$/2G (Fig.~3). The notations in Figs.~2 and 3 are
the same as in Fig.~1. Unlike $M_{\rm bh}$, the $M_{\rm nc}$ values for 
galaxies of early and late types form a unified sequence in all diagrams we considered. 

The dot-dashed line in Fig. 2b has the slope expected 
theoretically by models, in which the formation of SMBH and its final mass is linked with the mass of a dark halo: 
$M_{\rm bh} \sim M^{1.5}$, which corresponds to $M_{\rm bh}\sim 
V_{\textrm{max}}^{4.5}$~[18]. The shift of this line along the vertical 
axis is arbitrary. It is clear that the data for SMBHs agree poorly with the 
predicted slope. At the same time, this model slope is surprisingly close to the slope  fitted
for the NCs (dashed and dot-dashed lines in Fig.~2b). It is unclear, whether this agreement is a
coincidence, or if the gravitational potential of the halo determines 
the mass of a nuclear star cluster to a greater extent than the mass of
a central SMBH.

In Fig.~4, the masses $M_{\rm nc}$ and $M_{\rm bh}$ are compared with the
total masses of stellar population of galaxies, $M_\ast$. 
The latter was found from the total luminosity $L_B$ 
taking into account the mass/luminosity ratio 
$M_\ast/L_B$ as a function of color index  ($B{-}V$) [80]. The dependence we consider is apparent 
only for late-type galaxies, being more shallow for the NCs (Fig.~4a) than for SMBHs (Fig. 4b).
Using the whole set of galaxies of different types, including elliptical 
galaxies, Scott and Graham [12] also found $M_{\rm nc} (M_{*})$ relationship to be a
shallower than that for the SMBHs. 
In general, our results for the disk galaxies  are in good agreement 
with the results of [12].

It is worth noting, that the masses of both NCs and  SMBHs correlate more closely with dynamical masses $M_{25}$, which include both  stellar and dark-halo masses, than with pure baryonic stellar masses of galaxies $M_\ast$ (Figs.~3 and 4), despite the fact that the notion of dynamical mass is rather vague, because it contains the somewhat arbitrary adopted limiting radius $R_{25}$.

\begin{figure*}
\includegraphics[width=7.5cm,keepaspectratio]{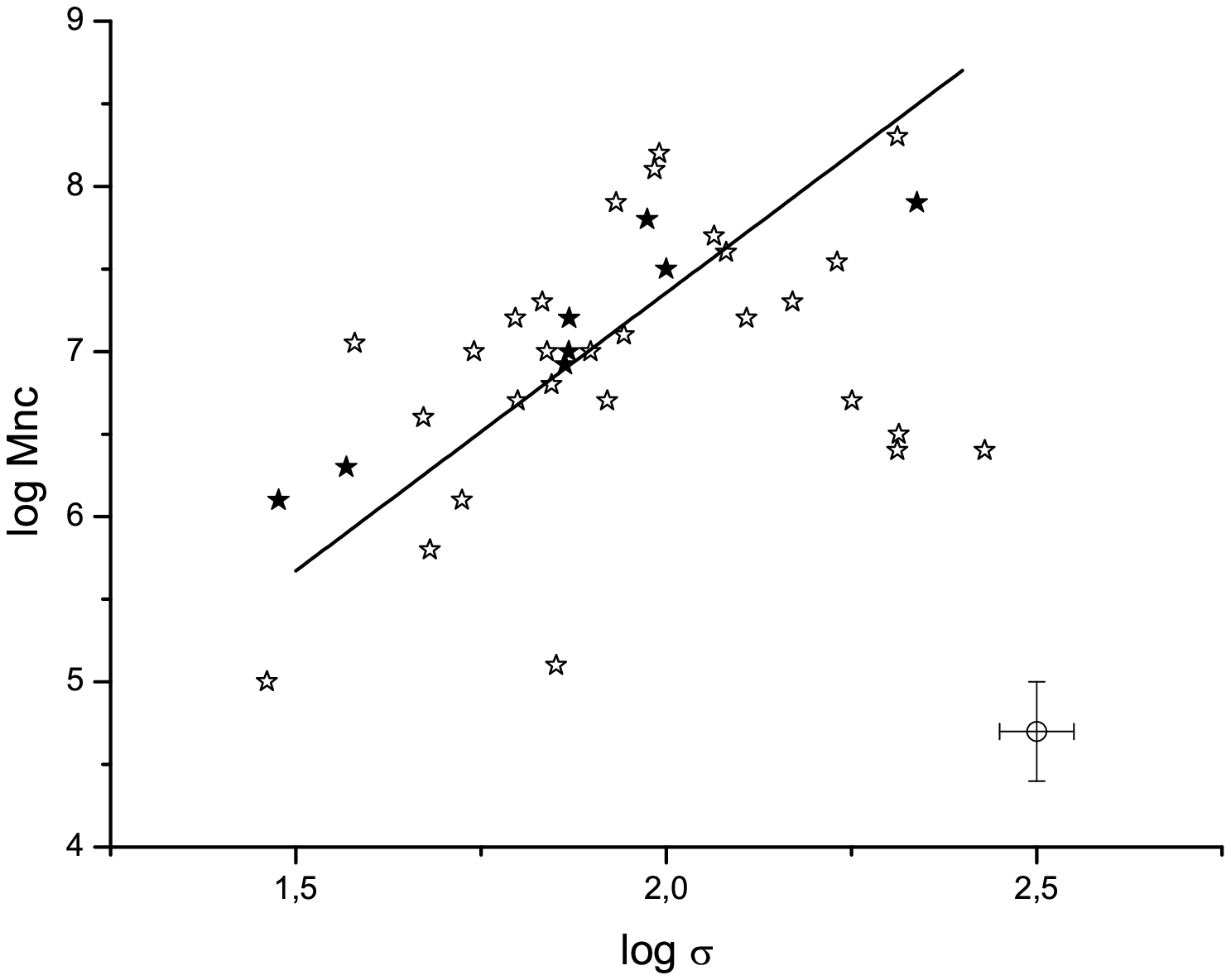}
\includegraphics[width=7.5cm,keepaspectratio]{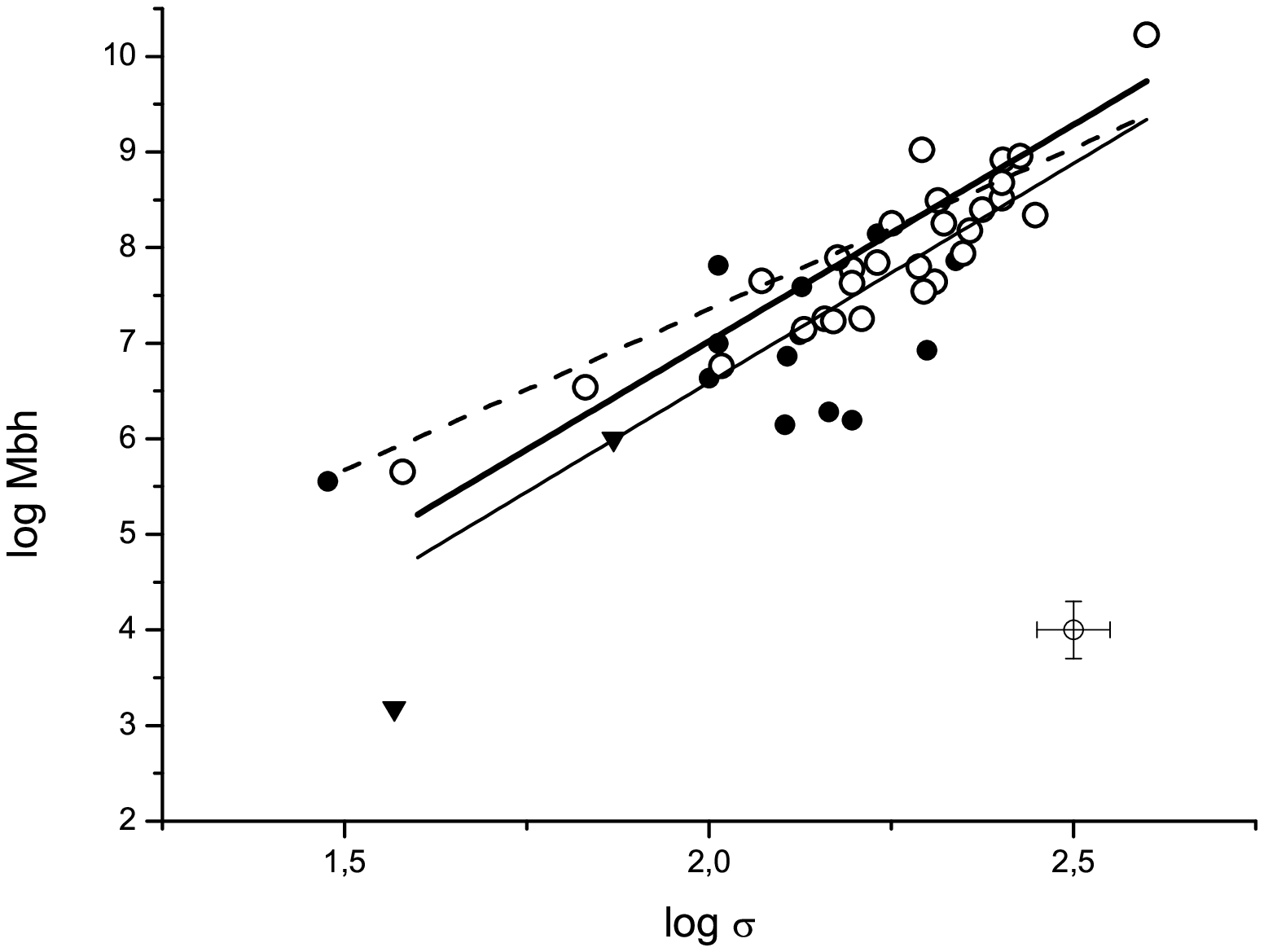}
\caption{ Relationship between (a) $M_{\rm nc}$ and (b) $M_{\rm bh}$ and
the central velocity dispersion $\sigma$.  The notation is the same 
as in Fig. 1. Two solid straight lines are taken from [81] (see the text).}
\label{fig5}
\end{figure*}


There is a somewhat different situation with the dependence of the masses 
of the central objects  on the central velocity dispersion $\sigma$. 
Though this relationship exists for both NCs  and SMBHs (Fig.~5), is is much 
weaker and  more gently sloped for the nuclear clusters than  for the black holes, which has been noted earlier by Scott and 
Graham [12]. The galaxies that most strongly deviate  towards higher 
velocity dispersions in Fig.~5a are lenticular galaxies with close $M_{\rm bh}$ values:  NGC~1023, VCC~1125 
= NGC~4452 and NGC~3585, whose disks are observed edge-on, and (at the 
bottom) the dwarf peculiar galaxy NGC~1705 with an outburst of star formation. It cannot be excluded that the high velocity disperions in 
these three galaxies result from the projection of different 
regions of the disk onto the line of sight, although other galaxies 
viewed edge-on do not show such an effect.  After the exclusion of these three 
galaxies, the correlation coefficient for the relation between $M_{\rm nc}$
and $\sigma_с$ turns out to be fairly high ($r = 0.65$). The dashed
line in Fig. 5b was fitted for the NCs and reproduced from Fig.~5a. 
The solid lines in Fig.~5b show the dependencies for SMBHs obtained by  
McConnell et al. [81]: the bold line corresponds to S0 and E galaxies
and the thin one --  to spiral galaxies. The lenticular galaxies in our
sample (small, hollow circles) lie on a common dependence with the spiral
galaxies, but are shifted by $\Delta \log M_{\rm bh}\approx 0.4$, on
average, relative to the dependence constucted in [81]  for elliptical
galaxies.

The parameters of linear regressions for galaxies with NCs and SMBHs,
corresponding to the half-angle between the direct and inverse regression lines,
are presented in Tables~2 and 3, together with their standard 
errors and correlation coefficients $r$.

\section{MASSES OF THE CENTRAL OBJECTS AND THE GALAXY COLOR INDICES}

Since the growth mechanisms for the NCs and SMBHs are directly or
indirectly connected with the evolution of stellar population, we  
consider below the relationship between their masses and  the colors of parent galaxies, 
using the ($B-V$) color index, corrected for absorption and 
reduced to a face-on orientation of the disk (according to [33]). For 
Mkn~279, which has a bright blue nucleus that affects the integrated 
color, we used the color index of the outer region, outside
the effective radius $R_e$ containing half the total luminosity.

Figure~6 shows a luminosity--color diagram for the galaxies. The overall 
sample of galaxies clearly splits into two groups separated by 
$(B{-}V)\approx 0.6-0.7$. Here we tentatively adopt $(B{-}V) = 0.65$ as the
boundary. These groups are well known as the red and blue sequences 
of galaxies in the bimodal distribution of color indices in the 
color--luminosity diagram (see, e.g.,~[82--85]). The blue group contains
galaxies with active star formation, whereas the red group 
is formed by passively evolving galaxies, in which star formation 
is very weak or absent.  This bimodal color distribution is a common feature of galaxies; it takes place
independently on the environment density of galaxies [85, 86] and corresponds
 fairly well to the separation into early and 
late morphological types (the hollow and filled symbols in Fig. 6).

\begin{table}[t!]
\caption{Main relationships between the NC mass and galaxy parameters, 
$\log M_{\rm nc}= A \log X + B$: standard errors $\Delta$ of the parameters 
$A$ and $B$ and the correlation coefficient $r$}
\begin{tabular}{lccc}
\hline
\multicolumn{1}{c}{$X$}& $A$& $B$& $r$\\
\hline
$V_{(2)}$&   \phantom{0}$3.4\pm 0.4$&   $-0.4\pm 0.8$&  0.74\\
$V_{max}$&   \phantom{0}$4.6\pm 0.4$&   $-3.0\pm 0.9$&  0.74\\
$M_{25}$ &   \phantom{0}$1.6\pm 0.1$&   $-9.7\pm 1.5$&  0.75\\
$M_*$    &  $1.55\pm 0.1$&   $-8.4\pm 1.2$&  0.71\\
$\Delta$ &   \phantom{0}$3.4\pm 0.5$&   \phantom{$-$}$0.6\pm 0.9$&   0.66\\
\hline
\end{tabular}
\end{table}

\begin{table}[t!]
\caption{Main relationships between the SMBH mass and the 
galaxy parameters, $\log M_{\rm bh}= A \log X + B$ (excluding galaxies 
with upper limits for $M_{\rm bh}$: standard errors $\Delta$ of the parameters 
$A$ and $B$ and the correlation coefficient $r$}
\begin{tabular}{lccc}
\hline
\multicolumn{1}{c}{$X$}& $A$& $B$& $r$\\
\hline
$V_{(2)}$&  $4.7\pm 0.7$&   \phantom{$-$0}$3.0\pm 1.5$&  0.70\\
$V_{max}$&  $6.2\pm 1.0$&   \phantom{0}$-6.8\pm 2.2$&   0.56\\
$M_{25}$&   $1.8\pm 0.3$&   $-12.0\pm 3.8$&  0.43\\
$M_*$   &   $1.6\pm 0.2$&   \phantom{0}$-9.7\pm 2.6$&  0.55\\
$\Delta$&   $4.4\pm 0.4$&   \phantom{0}$-2.1\pm 0.9$&  0.81\\
\hline
\end{tabular}
\end{table}

\begin{figure}
\includegraphics[width=7.5cm,keepaspectratio]{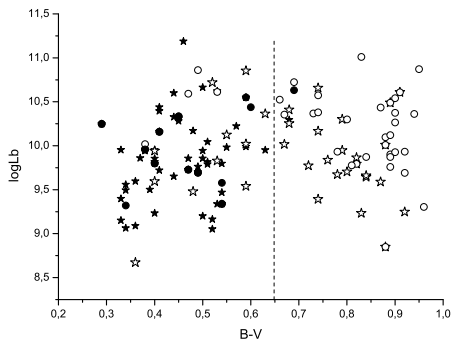}
\caption{$L_B$ --- ($B{-}V$) relation for the sample galaxies with known
estimates of $M_{\rm nc}$ (asterisks) and $M_{\rm bh}$ (circles). Hollow
symbols refer to early-type galaxies and filled symbols to late-type
galaxies. The vertical dashed line tentatively divides the galaxies 
into red and blue groups.}
\label{fig6}
\end{figure}


The bimodal nature of the galaxy color-index distribution indicates
that the  transition of galaxies from the blue to the red 
sequence must occur fairly quickly, at some stage of the evolution. 
Several hypothetical mechanisms for this transition have been considered 
(see the discussions of this problem by Gonsalves et al. [86], Mendez 
et al.~[87],  Gabor et al.~[88], di Matteo et al.~[89]).  Both internal and external 
mechanisms that can halt star formation, but do not destroy the stellar 
disk, are possible. If we exclude from consideration the rich  clusters of galaxies, where galaxies actively interact with the 
environment, an efficient mechanism that can halt
star formation in massive galaxies is the sweeping a gas out from the 
inner regions during a stage of high nuclear activity, 
which could be associated with galaxy mergers [89, 90].  Nuclear activity 
can also stop the accretion of gas from a halo onto a disk, which, 
in turn, inhibits the formation of stars [91].

\begin{figure*}
\includegraphics[width=7.5cm,keepaspectratio]{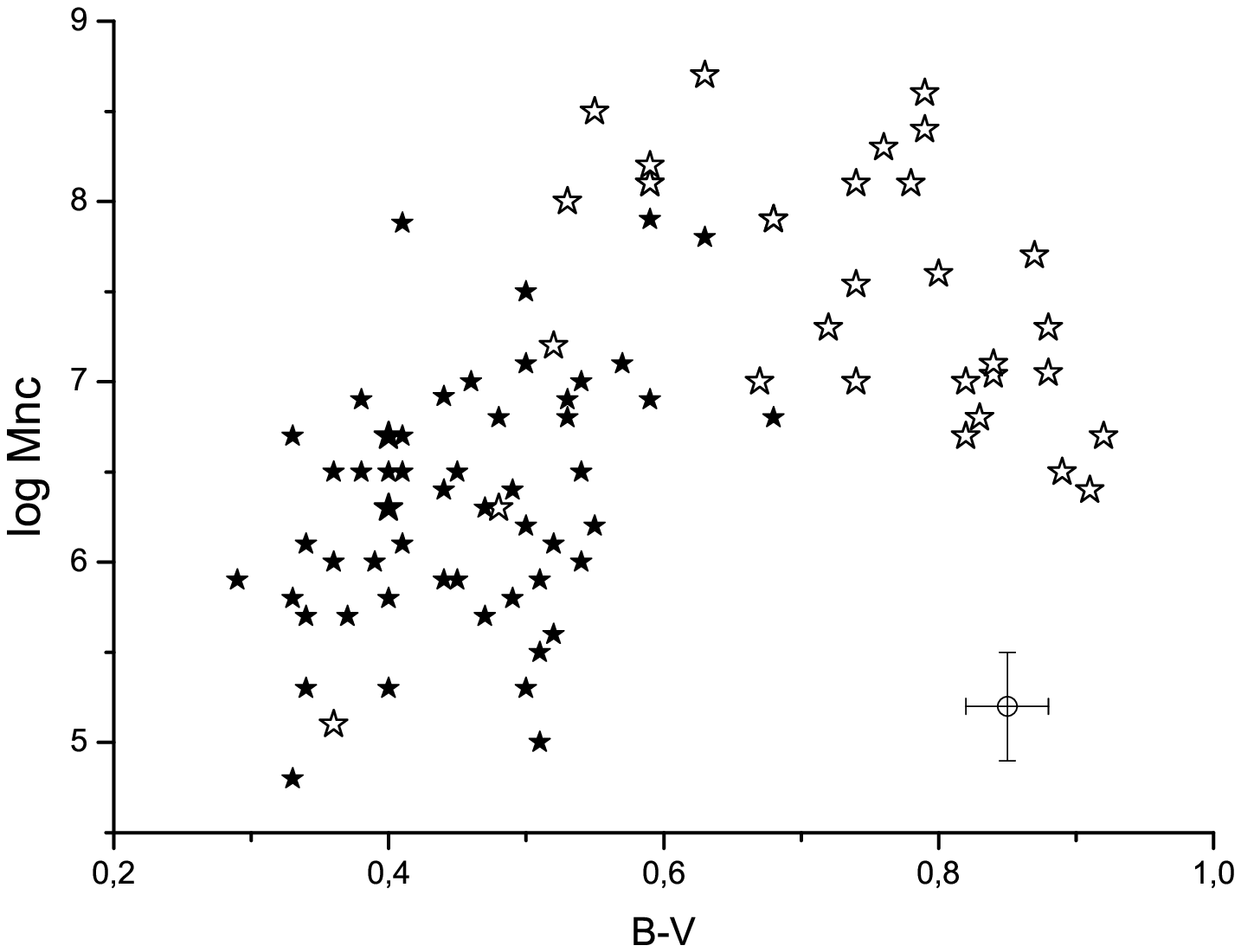}
\includegraphics[width=7.5cm,keepaspectratio]{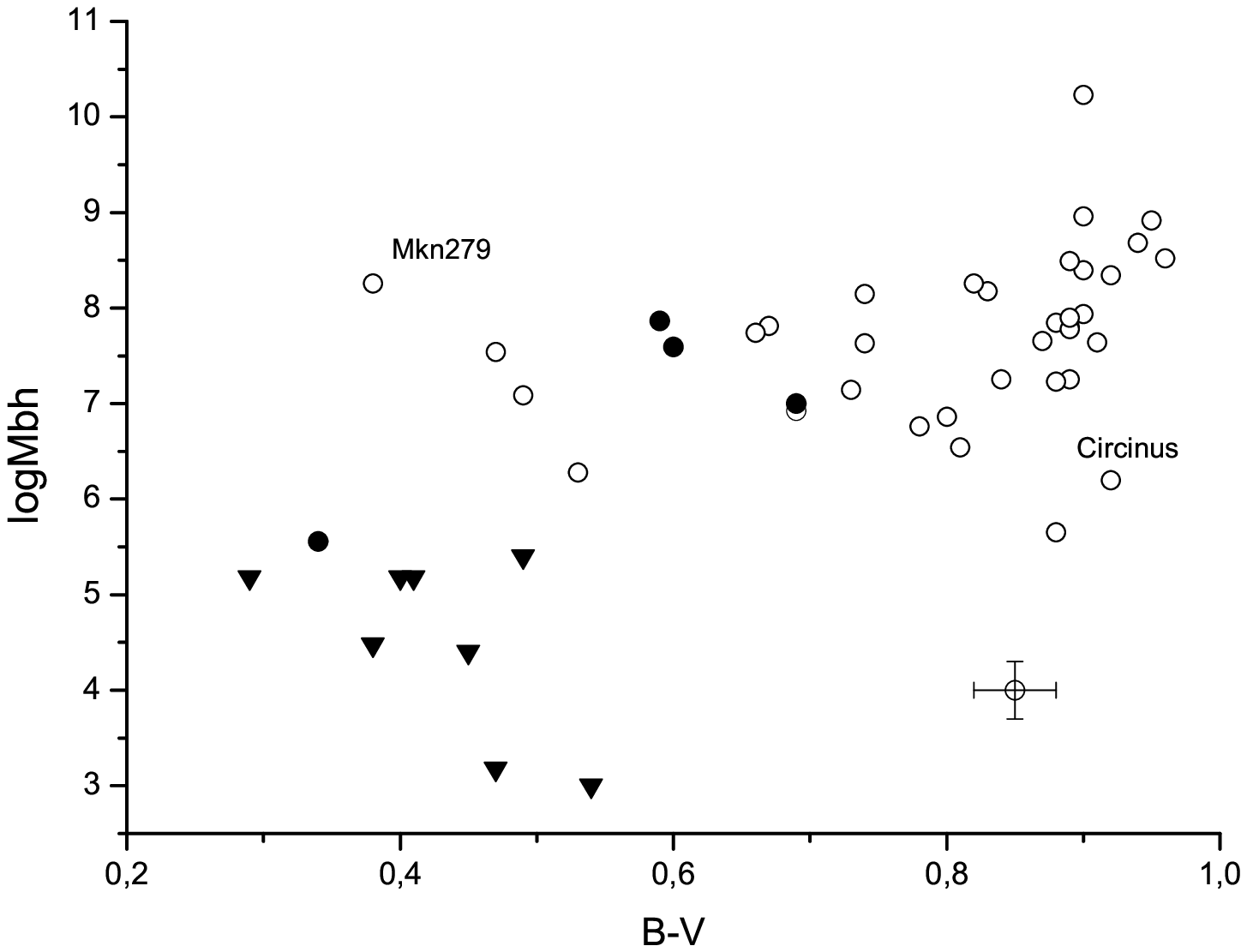}
\caption{ Galactic color --- central-object mass diagrams for (a) the NCs 
and (b) the SMBHs.  The hollow symbols correspond to the eary type galaxies.}
\label{fig7}
\end{figure*}


The estimated masses of the central objects -- NCs and SMBHs -- also dispart according to the
colors or morhological types of parent galaxies (Fig.~7). A red group contains more massive (in the mean)  nuclear clusters or black holes, whereas 
a blue group combines most of the low-mass central objects. 
At the same time, there is no dependence of $M_{\rm nc}$ or 
$M_{\rm bh}$ on the color index within each color group (for galaxies 
with massive NCs, there may even be an inverse dependence). An abrupt 
change in the relative mass $M_{\rm nc}$ between early and late-type galaxies was pointed out earlier by Erwin and Gadotti [9]. However, a considerable fraction of early-type galaxies in [9] are 
ellipticals, which have different star-formation histories, whereas here we consider the disky galaxies only.

Almost all galaxies, in which 
$M_{\rm bh}$ or $M_{\rm nc}$ does not exceed a few million $M_{\odot}$,
belong to the blue group. Thus,  the masses of NC or SMBH  are connected not only with the mass (density) of the central regions of galaxies, but also with the star formation in the entire disk.

\begin{figure*}
\includegraphics[width=7.5cm,keepaspectratio]{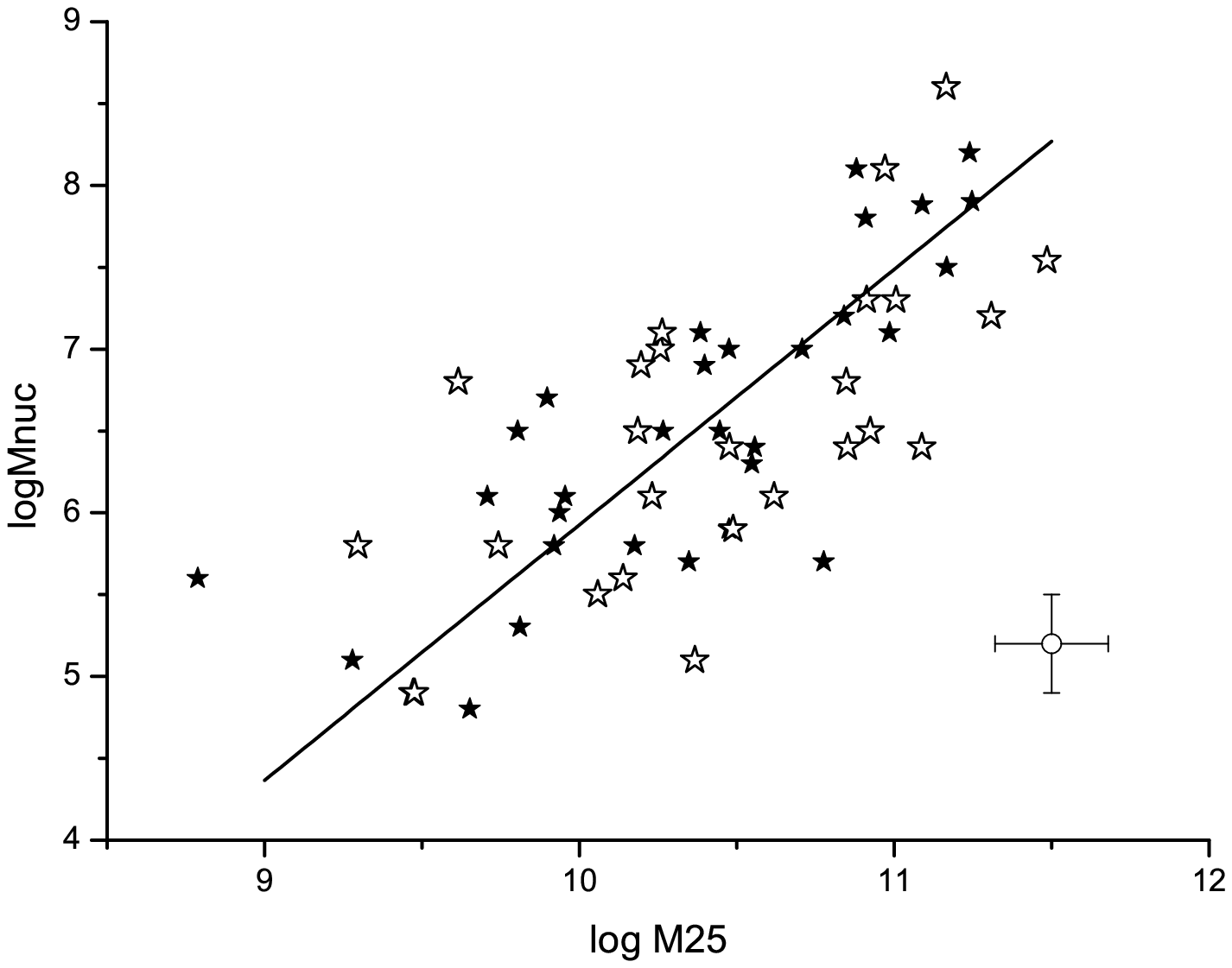}
\includegraphics[width=7.5cm,keepaspectratio]{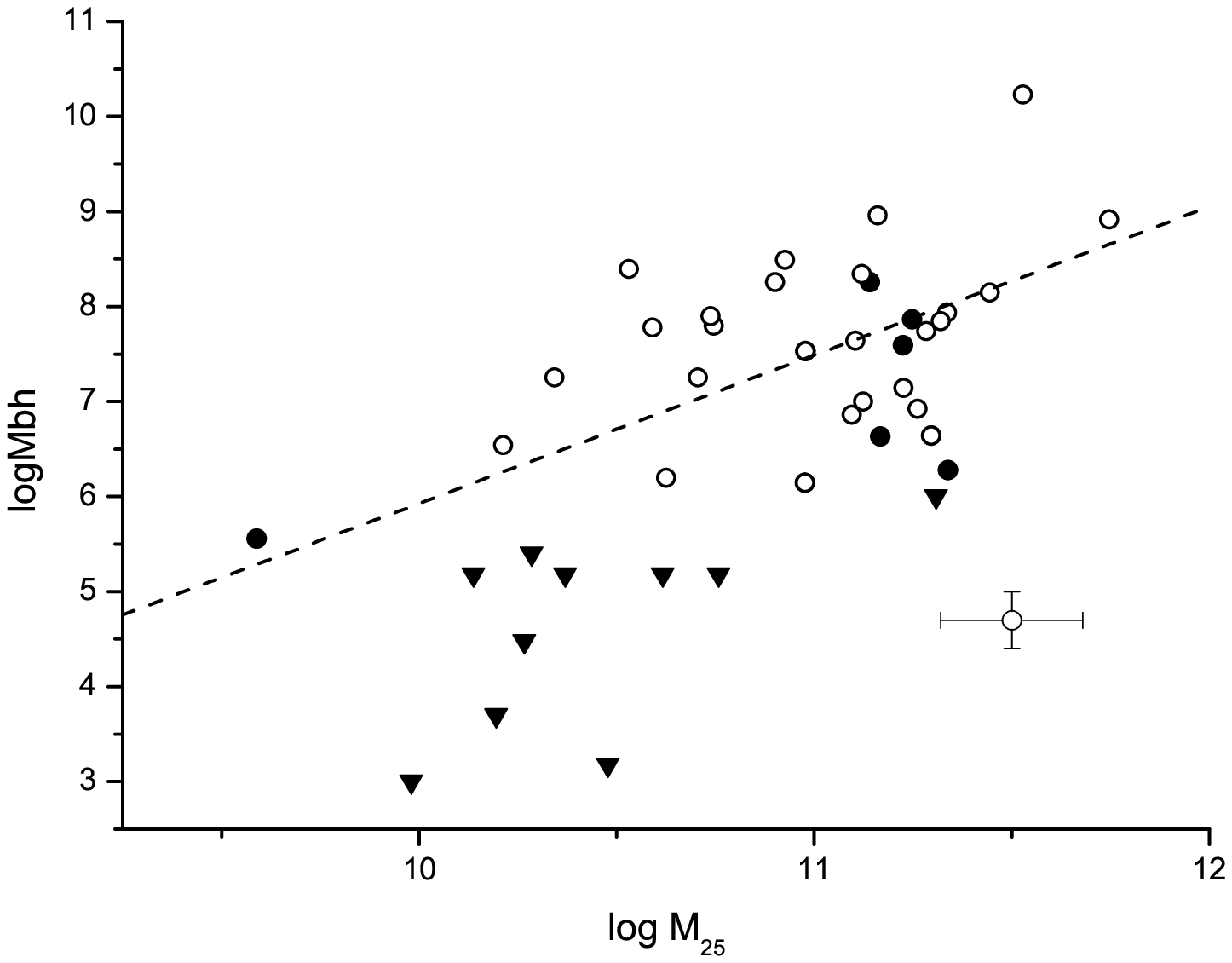}
\caption{ A version of Fig.~3 with the galaxies divided into blue and 
red groups (filled and hollow symbols). Both color groups form a single
dependence for the NCs in (a), however they are separated for the  SMBHs 
in (b).}
\label{fig8}
\end{figure*}


It is important that the ``red'' and ``blue''  galaxies
are not distinguished on the diagrams where the NC masses are compared with the properties of the host galaxies: their rotation velocity ($V_{(2)}$ or $V_{\textrm{max}}$), total mass, and mass of stellar population. In contrast,  SMBH masses $M_{\rm bh}$  are in most cases much higher for ``red'' galaxies than for ``blue'' ones, even for the analogous parameters of galactic disks. 
 This diference is most clearly
visible when we compare the masses of the central objects with the total masses  $M_{25}$ 
for galaxies of both color groups (Fig.~8). 
One may concludee that the  red-group galaxies usually possess much more massive central SMBHs than 
blue-group galaxies with the same rotation velocities (Fig.~8b), which 
is not true of the NCs (Fig.~8a). It agrees well with a scenario in which the decrease of  
star formation rate which makea a galaxy to be ``red`` is the result of high nuclear activity associated with already formed massive  SMBH at an early stage of the galaxy's evolution. The ejection
of a large amount of energy by an active nucleus leads to the compression
and throwing gas out of central region of a galaxy, or even from
the entire galaxy (see, e.g., [90--93]), or to the cessation of gas 
accretion onto the disk from the halo [91], which moves a galaxy into the passively 
evolving red sequence. Indications of intense gas outflows from the central 
parts of galaxies containing quasars are indeed observed [94]. Judging 
from the $M_{\rm bh}$ values for the ``red'' galaxies (Fig.~7b), the SMBH mass 
must exceed a few $\cdot 10^6 M_{\odot}$ to halt star formation, with a higher $M_{\rm bh}$ threshold 
probably required for more massive galaxies (Fig.~8b).

The impact of  active nucleus on 
the NCs is not entirely clear (see the discussion in [6]), however it is obvious
that it should halt  the rapid growth phase  of either SMBH or NC.  Since the red-group 
galaxies (mostly of earlier morphological types) cannot be distinguished 
in the $M_{\rm nc}-M_{25}$ diagrams (as well as in the other diagrams), the 
nuclear clusters inside these galaxies either managed to avoid the destruction during the active phase of nucleus, or they began their growth after this violent phase.

\section{CONCLUSIONS}

1. Using the available kinematic and photometric data for disk galaxies, we 
have confirmed the existence of correlations between the masses of the
central objects (nuclear clusters and supermassive black 
holes) and various kinematic parameters of the galaxies: the maximum 
rotation velocity, the angular velocity of the central region, the 
dynamical mass inside the optical radius, the central velocity dispersion,
and also the mass of the stellar population.  The correlations are tighter for 
NCs than for SMBHs, except for the correlation with the central velocity 
dispersion,  which  is closely connected with the SMBH mass. In all cases, the dependencies
for SMBH masses  ($M_{\rm bh}$) are steeper than for NC masses  ($M_{\rm nc}$), although this difference is determined  by the late-type galaxies only, whose central black holes, if present, have very low masses.

2. The NC masses  are probably more closely related with 
the galactic dark halos than are the SMBH masses. This is indicated by 
the tighter correlations between $M_{\rm nc}$ and the total mass of the
galaxy (which is the sum of the baryonic and dark masses) inside the 
optical radius, and also with the velocity $V_{\textrm{max}}$, which
characterizes the virial mass of the halo. In addition, the slope of the 
$M_{\rm nc} (V_{\textrm{max}})$ dependence turns out to be close to the 
corresponding relation for SMBHs in numerical models relating
the black hole' mass to the virial mass of the halo. This relationship between the NCs and the dark halo allows to suggest that the gravitational
potential well, created by the central density peak (cusp) of the dark 
matter (which may later disapppear, see, 
e.g.,~[95]),  may determine the growth condition of an NC in a young galaxy.

3. The black-hole masses $M_{\rm bh}$ are systematically higher  in 
lenticular galaxies than in spiral galaxies with similar values of
$V_{(2)}$ or $V_{\textrm{max}}$, $M_{25}$ or the mass of the stellar 
population (Figs.~1b, 2b, 3b, 4b). This indicates that the formation of 
at least a substantial fraction of S0 galaxies seems to be connected 
with the existence of massive central SMBHs in these galaxies.

4. The galaxies of our samples clearly divide into blue and red groups on the diagram ``Color index - total luminosity''.  The red group mostly contains S0--Sb 
galaxies with low star formation rate (Fig.~7). ``Red'' galaxies have much higher black-hole masses than the ``blue'' ones with similar host-galaxy masses or rotation velocities
(Figs.~1b, 2b, 3b, 4b, 8b). 

5.  The situation with the NCs is different: 
there is no statistically significant distinction between the galaxy
color groups in the ``NC mass --- Galaxy mass'' diagram (Fig.\,8).  
Irrespective of its color (although with a fairly large dispersion),
the mass of an NC is determined by parameters such as the disk rotation 
velocity, the dynamical mass inside the optical radius, and 
the total stellar mass of a galaxy. This agrees with a scenario in which the growth of the SMBHs was 
determined first and foremost by the physical conditions at the very 
center of the host galaxy, whereas the resulting masses of the NCs are
mostly related to the galaxies' integrated characteristics.

6. Our conclusion that the red-sequence galaxies tend to have  more 
massive central black holes agrees well with the idea that the division of galaxies 
 into two color groups may be a result of the partial or virtually 
total (as in the case of most S0 galaxies) loss of gas during a stage of high nuclear
activity caused by the existence of massive SMBH. The threshold value of 
$M_{\rm bh}$, providing the transition to the red group, is $10^6 - 10^7\ M_{\odot}$, and evidently depends on the total
mass of a galaxy. However, the occurrence of outbursts of nuclear 
activity need not imply the destruction of a nuclear cluster at the 
disk center, because its mass remains statistically tied with the properties of a parent galaxy. Otherwise the formation of central stellar cluster in a galaxy should begin after the phase of violent activity.

\section{ACKNOWLEGEMENTS}
We acknowledge the usage of the HyperLeda database (http://leda.univ-lyon1.fr).

\end{document}